\input harvmac
\newcount\figno
\figno=0
\def\fig#1#2#3{
\par\begingroup\parindent=0pt\leftskip=1cm\rightskip=1cm
\parindent=0pt
\baselineskip=11pt
\global\advance\figno by 1
\midinsert
\epsfxsize=#3
\centerline{\epsfbox{#2}}
\vskip 12pt
{\bf Fig. \the\figno:} #1\par
\endinsert\endgroup\par
}
\def\figlabel#1{\xdef#1{\the\figno}}
\def\encadremath#1{\vbox{\hrule\hbox{\vrule\kern8pt\vbox{\kern8pt
\hbox{$\displaystyle #1$}\kern8pt}
\kern8pt\vrule}\hrule}}

\overfullrule=0pt
\def\identity{{\rlap{1} \hskip 1.6pt \hbox{1}}}

\Title{TIFR-TH/00-47}
{\vbox{\centerline{ Magnetic Moments of Branes  and Giant Gravitons}}} 
\smallskip
\centerline{Sumit R.  Das\foot{das@theory.tifr.res.in},  Sandip P. Trivedi
 \foot{sandip@theory.tifr.res.in} and Sachindeo Vaidya
 \foot{sachin@theory.tifr.res.in}$^,$\foot{Address after Sept. 1st, 2000:
 Department of Physics, University of California, Davis CA 95616, U.S.A.}}
\smallskip
\centerline{  {\it Tata Institute of Fundamental Research,}}
\centerline{\it Homi Bhabha Road, Bombay 400 005, INDIA.}
\smallskip
\bigskip

\medskip

\noindent

We study the magnetic analogue of Myers' Dielectric
Effect and, in some cases, relate it to the blowing up of particles
into branes, first investigated by Greevy, Susskind and Toumbas.  We
show that $D0$ branes or gravitons in M theory, moving in a magnetic four-form
field strength background  expand into a non-commutative two sphere. Both
examples of constant magnetic field and non-constant fields in curved
backgrounds generated by branes are considered. We find, in all cases,
another solution, consisting of a two-brane wrapping a classical
two-sphere, which has all the quantum numbers of the $D0$ branes.
Motivated by this, we investigate the blowing up of gravitons into
branes in backgrounds different from $AdS_m \times S^n$.  We find the
phenomenon is quite general. In many cases with less or even no
supersymmetry we find a brane configuration which has the same quantum
numbers and the same energy as a massless particle in supergravity.

\Date{August  2000}

\newsec{Introduction and Summary}
We are getting increasing familiar with the idea that in string theory
particles grow in transverse size with increasing energy \ref\SZ{
L. Susskind, J. Math. Phys., {\bf 36} (1995) 6377-6396,
{\tt hep-th/9409089}; L. Susskind, ``{\it Particle Growth and 
BPS Saturated States}",{\tt hep-th/9511116}; T. Banks, W. Fischler,
S.H. Shenker and   L. Susskind, Phys. Rev. {\bf D 55} (1997) 5112-5128,
{\tt hep-th/9610043}}.  This idea
is supported by the string uncertainty principle 
\ref\uprin{T. Yoneya, {\it ``Duality and Indeterminacy Principle in String
Theory'' } in ``Wandering in the Fields'' eds K. Kawarabayashi and
A. Ukawa (World Scientific, 1987); T. Yoneya, Mod. Phys. Lett. A4 (1989)
1587; L. Susskind, Phys. Rev. D49 (1994) 6606.}.
and the IR/UV connection.
Another important and related development is that of
non-commutativity \ref\NC{A. Connes, M. Douglas and A. Schwarz,
J. High-Energy Phys. {\bf 9802} (1998) 003, {\tt hep-th/9711162}; 
M. Douglas and C. Hull, J. High-Energy Phys. {\bf 9802} (1998) 008, 
{\tt hep-th/9711165}}, \ref\SW{ N. Seiberg and E. Witten, J. High-Energy Phys. {\bf 9909}
(1999) 032, {\tt hep-th/9908142}.}, 
\ref\BS{D. Bigatti and L. Susskind, {\tt hep-th/9908056}} - the idea that 
space-time coordinates do not commute with each other.

An interesting example of the growth in size with energy was found
recently in \ref\GS{J. McGreevy, L. Susskind and N. Toumbas, JHEP 0006
(2000) 008, {\tt hep-th/0003075}.}.  These authors studied a graviton
\foot{More correctly we mean an appropriate supergravity
mode. Throughout this paper we will loosely refer to such modes as
gravitons.} in $AdS_m\times S^{p+2}$ which rotated on the $S^{p+2}$
and carried angular momentum. The graviton is a BPS state and has an
energy equal to its angular momentum. Somewhat surprisingly, \GS\
showed that the same BPS relation is satisfied by an expanded brane
configuration. For large angular momenta, \GS\ argued, the
graviton blows up into the expanded brane whose size increases with
increasing angular momentum for $p \geq 2$. Since the size of the
expanded brane is bounded by the radius of the $S^{p+2}$ there is a
maximum bound on the angular momentum; this agrees with the stringy
exclusion principle \ref\exclusion{J. Maldacena and A. Strominger, JHEP 9812
(1998) 005, {\tt hep-th/9804085}}.

The phenomenon described above is quite similar to Myers' Dielectric effect
\ref\M{R. C. Myers, ``{\it Dielectric Branes}", JHEP 9912 022 (1999),
{\tt hep-th/9910053}.}. It was found in \M\ that due to the non-Abelian
nature of their world volume theory $N$ $D0$-branes placed in an
electric $RR$ four-form field strength expand into a noncommutative
two-sphere. There is another solution in the theory consisting of a
$D2$-brane which wraps the corresponding classical two-sphere.  The
$D2$-brane carries $N$ units of $U(1)$ magnetic field in its world
volume and has exactly the same quantum numbers as the $D0$-brane
configuration.  

This paper explores the relation between \GS\ and \M\
and extends the analysis of \GS\ to more general settings.

We begin by demonstrating the magnetic analogue of the Dielectric
Effect (we will refer to this as the Magnetic Moment effect below). 
 A simple controlled setting is provided by a constant four
form  magnetic field, $F^{4}_{1234}$ in Type II theory. One
finds that $D0$-branes, when moving in the magnetic field, blow up
into a non-commutative two-sphere. The size of this sphere increases
with increasing momentum. We also find a minimal energy $D2$ brane
solution which has the same momentum and wraps the corresponding
classical two-sphere. 
Both the puffed $D0$ branes and 
$D2$-brane carry a magnetic  dipole moment with respect to $F^4$.
Variants  of the Dielectric
effect were studied  in 
\ref\TV{S. P. Trivedi and S. Vaidya,
``{\it Fuzzy Cosets and their Gravity Duals}",
{\tt hep-th/0007011}.}
leading to fuzzy $S^2 \times S^2,~CP^2$ and ${SU(3) \over
U(1) \times U(1)}$. Their magnetic analogues are also discussed.
The resulting configurations can carry dipole or quadropole magnetic moments.

$D0$ branes are gravitons in M-theory. This leads us to consider a
situation where $F^{4}_{123M}$ is non-zero and the graviton moves
along the $M$ direction.  In the Type IIA limit this reduces to static
$D0$ branes in a constant $NS$  $H_{123}$ field. Once again, following, \M,
we show that the $D0$-branes "puff up'' into a non-commutative
two-sphere.  The size of this sphere grows with increasing momentum in
the $M$ direction.  There is an alternative description of this configuration in terms of a 
 two-brane wrapping a two-sphere.

Once the Magnetic Moment  Effect is understood for constant magnetic field one
can study it in more complicated examples. Towards the end of the
paper we study $D0$-branes moving in the background of a
$D4$-brane. In this case the background geometry is curved and the
four-form field strength threads a four-sphere (of varying radius).
We show that once again rotating $D0$-branes expand into a
two-sphere. Moreover the resulting configuration has exactly the same
energy as if the $D0$ branes were executing only center of mass motion
with no relative displacement.  We also find a $D2$-brane solution
with $U(1)$ flux on its worldvolume, 
which has exactly the same quantum numbers and in fact (in the
appropriate limit) the same energy.

The understanding of the Magnetic Moment   Effect prompts us to extend the
analysis of \GS\ to more general settings.  In a sense the results of
\GS\ are surprising since one finds that expanded brane configurations
- which are normally thought to be heavy objects - can have the same
energy as massless particles.  This happens because the coupling to
the magnetic field threading the $(p+2)$ sphere precisely cancel the
effects of the brane tension.  At first sight this appears to be a
very special feature of a particular kind of motion in $AdS_m\times
S^{p+2}$ space-times.  It was shown in
\ref\djm{S.R. Das, A. Jevicki and S.D. Mathur, ``{\it Giant Gravitons,
BPS bounds and noncommutativity}", {\tt hep-th/0008088}} that the energy for
arbitrary brane motions obeys a BPS {\it bound} \foot{ We use the phrase BPS
bound in the original sense of term, viz. the fact that the energy
is bounded from below by a conserved charge} and the special motion
considered in \GS\ (i.e. motion without change of size and without
oscillations on the brane ) saturates the bound. Furthermore in a
supersymmetric theory these motions have been shown to preserve half of the
supersymmetries of the background \ref\MG{M. Grisaru, R. Myers
and O. Tafjord, {\tt hep-th/0008015}; A. Hashimoto, S. Hirano and
N. Itzhaki, {\tt hep-th/0008016}.}  so that the BPS bound follows from
supersymmetry. The derivation of the BPS bound in \djm\ follows
from a delicate cancellation which depends on detailed form of
the background and one might wonder whether the result has any level
of universality.

We find that gravitons can turn into expanded branes in other
spacetimes as well. Contrary to the expectations mentioned above it
turns out that in several cases, including spacetimes with no
supersymmetry at all, the brane solution has the same energy as the
graviton.  More specifically, we study gravitons in various extremal
and non-extremal $(6-p)$-brane backgrounds.  These backgrounds preserve
a $SO(p+2)$ subgroup of the $R$-symmetry group and the gravitons carry
$SO(p+2)$ charge. In the near horizon limit, for both the extremal
and {\it non-extremal} cases, we find that a $p$-brane configuration,
which is the  solution of least energy for a given angular momentum, has
exactly the same energy  and motion as a graviton.
 Moving away from the near horizon limit the
energies do not agree anymore.  The extremal background geometry, for
$p\ne 3$ has half the supersymmetries as the $AdS$ cases studied in
\GS.  As best as we can tell, the $Dp$-brane configuration moving in
this background does not preserve any of them
\foot{In these examples, there is generically no BPS bound, except for
$p=3$.  We work in Poincare coordinates. In such coordinates a
graviton in the near horizon region with some angular momentum is not
static but would fall into the ``center''. Only for $p=3$ one can go to a
different coordinate system - global coordinates in $AdS_5$.}.
The non-extremal geometry clearly breaks all
supersymmetries. Similar results also hold for the non-extremal
$M2$ and $M5$ branes. 

Our considerations also 
apply to  the extremal and
non-extremal geometries of five and four dimensional black holes in
string theory obtained by compactification on $T^5$ and $T^6$
respectively. Once again one finds 
brane configurations with the same energy as gravitons.
We should add though that the interpretation of the brane  configurations
in these cases as gravitons is not so clear. 

It is worth emphasising that expanded $p$-brane configurations
with the same energy as massless particles can be found 
only if an important condition is satisfied.
The  $p$-brane
in the course of its motion sweeps out a $S^p\times S^1$ surface.
The metric seen by the  
$p$-brane is related to the string metric by a conformal rescaling.
For $Dp$-branes the  condition says that the 
volume of this surface in the $p$-brane metric  must equal
(in appropriate string units) the number of $(6-p)$ branes. 
In the $M$ theory cases a similar condition must  hold without any conformal rescaling.
Unfortunately, the significance of this condition is unclear to us at the 
moment. It is worth pointing out though that in all cases the  $p$-brane which acts 
like a giant graviton is  the magnetic dual of the brane  which produce the geometry. 

One can be justified in claiming that gravitons turn into
expanded branes only if the two descriptions, of graviton 
and expanded brane are not simultaneously valid.
In section 5 we show that this
is indeed true.  The graviton description is controlled, in the
supergravity approximation, for small angular momenta, when higher
derivative corrections to the supergravity Lagrangian can be
neglected. In contrast the expanded brane description is valid for
large angular momenta, when the size of the brane is large , and  its
curvature is small, so that the Born Infeld action which neglects higher
derivative terms can be used. We establish this
 both for the $AdS$ case studied in
\GS and for the general $p$-brane backgrounds considered in this paper.

Our study is incomplete in some important respects.

The Magnetic Moment Effect discussed here provides a quantitative 
understanding of expanded branes in only one case:
$D0$ branes rotating in the $D4$ brane background,
which expand  into  a $D2$ brane
with magnetic flux, as was mentioned above. Such an understanding 
 for expanded branes without a 
world volume magentic flux is missing though. In particular we do not
have a good understanding of  
the blowing up of gravitons in  $AdS_m \times S^{p+2}$ or $Dp$-brane  backgrounds.
In the $AdS_m \times S^{p+2}$ case, 
for $(m,n)=(4,7)$,  and, $(7,4)$ though, the Magnetic effect does
provide at least a qualitative  understanding. 
In these cases if we are cavalier and regard one of
directions of the $S^n$ to be the M theory direction, gravitons
moving in this direction are $D0$ branes. From the IIA viewpoint
these $D0$ branes are in a background of a NS 3-form field strength
for $(m,n) = (7,4)$ and a RR 6-form field strength for
$(m,n) = (4,7)$. In the former case one would expect the $D0$ branes
to blow up into fuzzy $2$-spheres as shown in this paper.
In the latter case, the couplings
indicate that there could be solutions with noncommutative 4-spheres,
though this is not quite understood yet.

Another  important question which remains is to understand the blowing up
of gravitons or $D0$ branes 
in the dual holographic gauge theory.  For example, the
$D0$-branes mentioned above in the $D4$ brane background are Yang
Mills instantons in the boundary theory. One can ask : what is the
holographic description of their blowing up into $D2$ branes\foot{ The
analogous question for branes expanding in $AdS$ space was studied in
\MG.
 We should mention that in this  paper we will use Poincare coordinates as 
opposed to
global coordinates.  The  boundary theory in this case lives in flat
space and does not have an $R\phi^2$ term coupling. Correspondingly
there are no finite energy  states in the bulk with branes extended in the $AdS$
directions.}  ? More generally one can ask the same question about
gravitons and other massless states in the gravity theory.  The fact
that the giant graviton phenomenon is more general may be important in
understanding the structure of spacetime at short distance scales.  It
has been also argued that the stringy exclusion principle and some of its
manifestations \ref\JM{
O. Lunin and  S. Mathur, {\tt  hep-th/0006196} ; 
A. Jevicki, M. Mihailescu and S. Ramgoolam, {\tt hep-th/0006239}
. }
means that the dual supergravity should live on a noncommutative space-time,
e.g. quantum deformations of $AdS \times S$
\ref\ramjev{A. Jevicki and S. Ramgoolam, JHEP 9904 (1999) 032,
{\tt hep-th/9902059}; P. Ho, S. Ramgoolam and R. Tatar, Nucl. Phys.
B 573 (2000) 364, {\tt hep-th/9907145}.}
, for a related effect in matrix theory see \ref\BV{M. Berkooz and H. Verlinde, 
``{Matrix Theory, AdS/CFT and Higgs-Coulomb Equivalence.}", {\tt hep-th 9907100}.}.
It has been suspected that the dynamics of giant
gravitons, in particular the upper bound on the angular momentum for
special class of states, point to such a noncommutativity - a connection
which has been explored in \djm\ and \ref\holi{P. Ho and M. Li,
  {\tt hep-th/0004072}}. A holographic description would perhaps relate
this spacetime Non-commutativity to Non-commutativity in the
boundary  theory.

Finally, the process by which 
$D0$-branes or gravitons turn into the corresponding expanded brane,
seems related to the decay of brane -antibrane pairs into
lower dimensional branes \ref\SEN{A. Sen, "{\it Non-BPS States and Branes in 
String Theory}'', {\tt hep-th/9904207}.}. 
Investigating the connection in more detail would be worthwhile.

\newsec{The Magnetic Moment  Effect}
\subsec{The Electric Myers Effect}

We begin by briefly recalling the electric Myers effect \M\  (the fuzzy sphere
in matrix theory was  considered in \ref\taylor{D. Kabat and W. Taylor, 
Adv.Theor.Math.Phys. {\bf 2},
181-206 (1998), {\tt hep-th/9711078}.}) .
Consider $D0$-branes in a transverse electric four form field strength 
background:
\eqn\three{F^{(4)}_{0ijk}= \cases{-F \epsilon_{ijk}, &
 for $ i,j,k \in \{ 1,2,3 \}$; \cr 0 & otherwise \cr}}
 where $F$ is a  constant.
For a static configuration, the  resulting $D0$ brane Energy  is given by
\eqn\act{E=T_0 N -  { T_0 \over 4 \lambda^2}
\sum_{ab} Tr([X^a,X^b]^2) - i {T_0 \over 3 \lambda} 
Tr(X^iX^jX^k)F^{(4)}_{0ijk}.}
The last term in \act\ was discussed in \ref\taylortwo{W. Taylor and 
M. Van Raamsdonk, {\tt hep-th/9904095}; W. Taylor and M. Van Raamsdonk, 
{\tt hep-th/9910052}.} and \M.
In our conventions
\eqn\consdef{\eqalign{
T_p=&{2 \pi \over g_s (2 \pi \l_s)^{p+1}}, \cr
\lambda=&2 \pi \l_s^2. }}

For static configurations, one can show that the energy is minimized by 
setting,
\eqn\elesol{X^i={\lambda F \over 2} J^i, i=\{1,2,3 \} }
where $J^i$ denote $N$ dimensional representation of $SU(2)$,
with the remaining coordinates being proportional to the identity matrix.
The solution \elesol\  is  a fuzzy two-sphere. Choosing $J^{i}$ to be in the 
$N$ dimensional irreducible representation of $SU(2)$ gives a radius
and energy for the solution \elesol\ :
\eqn\charele{\eqalign{R=&{ \lambda \over 2 } F \sqrt{N^2-1\over 4} \simeq 
                         {\lambda \over 4} F N    \cr
                      E=&T_0N  -{T_0 \over 96 } \lambda^2 F^4 N {N^2-1 \over 4} 
                         \simeq T_0N - {T_0 \over 384} \lambda^2 F^4 N^3, }}
where the two approximate equalities relate to the $N \gg 1$ limit.
We also note that the resulting configuration 
carries  a dipole moment with respect to the four form field strength. 

Another configuration with the $N$ units of $D0$ brane charge and the 
same dipole moment with respect to $F^{(4)}$ can be constructed in terms of 
one $D2$-brane wrapping a sphere in the $X^1,X^2,X^3$ directions. 
The $D2$-brane carries $N$ units of $U(1)$ world volume magnetic flux.
The energy  for such a static  brane, which follows from the 
Dirac Born Infeld action and the Cherns Simons terms, (for a review see 
\ref\TS{A. A. Tseytlin, ``{\it Born-Infeld Action, Supersymmetry and 
String Theory}'', {\tt hep-th/9908105}, and references therein. } )
  is given by
\eqn\eleex{E=4 \pi T_2  \sqrt{r^4 + {N^2 \over 4} \lambda^2}
- {4 \pi \over 3} T_2 F r^3  }
where $r$ denotes the radius of the two-sphere and we have substituted for 
$F^{(4)}$ from \three.
Notice, that the energy does not have a global minimum and 
goes to $-\infty$ as $r \rightarrow \infty$.
This indicates an instability for the two-brane to grow very  big.

There can,  however, be a local minimum for a suitable range of parameters.
When 
\eqn\conda{r^2 \ll N \lambda,}
\eleex\ can be expanded as:
\eqn\expaact{E \simeq 2  \pi T_2 \lambda N + {4 \pi T_2 \over \lambda} {r^4 \over N}
-{4 \pi \over 3} T_2 r^3 F.}
This gives a minimum at a radius
\eqn\elecr{R={\lambda \over 4} F  N,}
and an energy equal to \charele.
Consistency with \conda\ imposes the condition:
\eqn\condcream{N \lambda F^2 \ll 1.}
More generally the full energy \eleex\ needs to be minimised. 
One can show that a local  minimum only exists if
\eqn\condb{F^2 < { 4 \over N \lambda}.}

A few comments are  worth making at this point.

The expression for the energy \act\ is an approximation;
in general there are additional terms involving higher powers
of the transverse coordinates. This
approximation is justified only if the radius of the two-sphere is
small compared to the string scale (the masses of the "W" bosons are
then small in string units).  In contrast for the $D2$ brane, the
coordinates along the two-sphere lie along the world volume. The DBI
action which gives rise to the energy, \eleex, is a good approximation
when the size of the two-sphere is big compared to the string
scale. In this limit higher derivative terms - that is "acceleration
terms" - can be neglected.  Thus we see that, in general,  the two descriptions, in
terms of the puffed up $D0$ branes and the wrapped $D2$ brane,   are
valid in different regions of parameter space.  
Not surprisingly,  the energy and radius of the 
fuzzy sphere derived from both descriptions do not agree. 

Agreement is obtained in the limit \conda\ though. In fact in this limit,
not only does the radius and the energy of the $D2$ brane agree with that of \charele,
but  each  terms of the expansion \expaact, agrees with \act.
A little thought, along the lines of \SW,  shows that this agreement is  to be expected. 
The important point is that the 
$D0$ brane desription is valid when  the two-sphere has  a radius,  measured in the 
{\it closed string metric} , which is  small compared to the string scale, while the $D2$ brane action is valid when 
the radius,  in the {\it open string metric}, is big compared to the string scale. 
When \conda\  is true, 
an argument similar to that in \SW , shows that both requirements are  met simultaeneously 
(for large $N$).

Finally, we have neglected the curvature of the spacetime due to the 
$F^{(4)}$ field strength. Strictly speaking, this back reaction needs to be incorporated \foot{
Note, in our conventions, \act\ the action goes like $S=-\int {1 \over g_s^2} (F^{4})^2 + 
\cdots $.} .
Our neglect can be justified if the theory under consideration is a boundary hologram,
or perhaps, if there are other sources, besides $F^{(4)}$, cleverly turned on to keep the metric
 flat and dilaton constant. Later on, in the context of the   magnetic effect, we will consider 
examples where the back reaction is included.

\subsec{The Magnetic Moment Effect}
 
 Consider a four form 
background of the form:
\eqn\mago{F^{(4)}_{ijkl}= \cases{-F \epsilon_{ijkl}, &
 for $ i,j,k,l \in \{ 1,2,3,4 \}$; \cr 0 & otherwise \cr}}
The background preserves a $SO(4) \times SO(5)$ symmetry. 
The resulting Lagrangian is now 
\eqn\actmag{L=-T_0 N + {T_0 \over 2} Tr({\dot X^i})^2 + 
{ T_0 \over 4 \lambda^2}
\sum_{ab} Tr([X^a,X^b]^2) + i {T_0 \over 3 \lambda}F^4_{ijkl}
Tr[X^iX^jX^k({\dot X^l})] }
The  derivative above is with respect to time. The last term is linear in the velocity
as is usual in a coupling to the magnetic field, it was also considered in \taylortwo. 

One can show that the equations of motion which follow from 
\actmag\ and \mago\ can be solved by
\eqn\magsol{\eqalign{X^4=&v X^0 \identity    \cr
            X^i=&{\lambda \over 2} F v J^i,   i =\{ 1,2,3 \}.  }}
with all other $X^a$'s being constant and proportional to the identity
matrix.
Other solutions to \actmag\ and \mago\ can be obtained by performing a $SO(4)$
rotation on the four coordinates.

The $D0$ branes have thus expanded into a non-commutative two-sphere
in the directions transverse to the velocity. The radius of this
two-sphere depends linearly on the velocity and the four form field
strength.  If we choose the $J^i$ matrices in \magsol\ to be in the
$N$ dimensional irreducible representation the radius is
\eqn\radmag{R={\lambda \over 2} F v \sqrt{{(N^2-1) \over 4}} 
\sim {\lambda \over 4} F
 v N, } where the approximate equality is valid for large $N$.  The
 energy for this case (when $N \gg 1 $) is
\eqn\enemag{E=N T_0 +  {T_0 \over 2} N v^2 - {T_0 \over 384 }  
\lambda^2 F^4 N^3}
Note that if we express the radius in terms of
the momentum of $D0$ branes,
which is
\eqn\momen{P_4 = NT_0 v+{2\over 3} {F r^3 T_0 \over \lambda}}
in our approximation, $N$ drops out and one has
\eqn\radmaga{ R = {\lambda F \over 4 T_0}(P_4 - {2 \over 3}{F r^3 T_0 \over \lambda}). }

For reasons mentioned in the   the electric 
case, the action \actmag\ is a good approximation when the radius
\radmag\ (measured in the closed string metric) is small in string units. 
We should also mention that choosing a reducible representation 
in \magsol\ gives rise to more than one fuzzy sphere, in general of 
different radii.

Next, 
consider a  $D2$-brane wrapping a sphere of radius $r$ in the $1,2,3$ directions
 and moving in the 
$X_4$ direction. Its action is 
 given by 
\eqn\acmagdtwo{S=\int dt [ -4 \pi T_2 \sqrt{1-({\dot r})^2-({\dot X^4})^2}
\sqrt{r^4+{N^2\lambda^2 \over 4}}
+{4 \pi \over 3} T_2 r^3 F {\dot X^4} ]. }
The terms above within the square root  arise from the Born Infeld action
 while the  last term comes from the Cherns Simon action which has the   form
\eqn\csform{S_{CS}=T_p\int C^{p+1},}
for a $p$-brane.
Assuming the motion is non-relativistic and that $r^2 \ll N \lambda$,
one has,
\eqn\apprmagdtwo{S\simeq \int dt [-2\pi T_2 \lambda N + 2 \pi T_2 N \lambda 
({\dot r})^2
+2 \pi T_2 N \lambda  ({\dot X^4})^2 -{4 \pi T_2 \over N \lambda}r^4+ 
{4 \pi \over 3} T_2 r^3 
{\dot X^4}
F].}
Putting in the ansatz for a non-commutative two-sphere of radius $r$
in \actmag, one finds that \actmag\ and \apprmagdtwo\ agree term by
term with ${\dot X^4}$ being identified with $v$. 
Thus setting, ${\dot r}$ equal to zero and minimizing the action
with respect to $r$ gives a radius and an energy from \apprmagdtwo\
which agrees with \radmag\ and \enemag.

However, in the more general case when the motion is relativistic 
or $r^2 \ge N \lambda$,
one needs to work with the full action, \acmagdtwo.
It is useful then to discuss the dynamics in terms of the Hamiltonian.
We have:
\eqn\hammag{\eqalign{P_r={\partial L \over \partial {\dot r}}&=4\pi T_2
\sqrt{r^4+{N^2 \lambda^2 \over 4}} {{\dot r} \over
\sqrt{1- ({\dot r})^2 -({\dot X^4})^2} }   \cr
P_4={\partial L \over \partial {\dot X^4}}&=4\pi T_2
\sqrt{r^4+{N^2 \lambda^2 \over 4}}{ {\dot X^4} \over 
\sqrt{1- ({\dot r})^2 -({\dot X^4})^2} }+{4\pi \over 3}T_2r^3 F }}
The Hamiltonian is 
\eqn\hamtwo{H=[(4 \pi T_2)^2 (r^4 + {N^2 \lambda^2 \over 4})+P_r^2
+(P_4-{4 \pi \over 3}T_2 r^3 F)^2]^{1/2}}
$P_4$ is a constant of motion. 
Restricting to motion with constant 
radial size, we set $P_r=0$ and minimize $H$ with respect to $r$.
It is easy to check that apart from the trivial solution $r = 0$ there
is always a single real solution of this equation with nonzero $r$. 
In general this gives an energy and a radius $r$ different 
from \radmag\  and \enemag. 

Let us end with  a few  comments.
First, in the magnetic case  there is a one parameter
family of   solutions depending  on $v$ in \enemag\ or $P_4$ in \hamtwo.
The transverse size of the sphere depends on this parameter. 
In fact, from \hamtwo\ it is clear that the equilibrium 
radius depends on $P_4$ only and not
on $N$, consistent with \radmaga.
Second, one could have guessed the form of the solution for the magnetic case, 
from electric one by  performing a boost. But strictly speaking
one cannot go from the purely electric to purely magnetic case by a boost.
Third, as in the electric case we have neglected the backreaction on the 
metric due the $RR$ field strength. We will comment, briefly, on this issue in the next section.  
Fourth, the resulting fuzzy two-sphere carries  an electric dipole moment which  
couples to the electric  four-form  field strength. It also has a magnetic dipole moment
which couples to the magnetic four-form field strength.
The induced magnetic dipole moment results in lowering the energy of the configuration
(for fixed momentum  $P_4$), as seen in the last term in \enemag\ or \hamtwo\. 
In this sense the
system behaves like a paramagnet. The decrease in energy - which goes like $N^3$  in \enemag\ - 
can be significant for large $N$

Finally, generlisations of the Dielectric effect which yield fuzzy 
$S^2 \times S^2$, $CP^2$ and $SU(3)/(U(1) \times U(1))$ were studied in 
\TV. The corresponding magnetic generalisation are straightforward.
Replace \mago\ by 
\eqn\fuzzf{F^{(4)}_{ijk9}= -F f_{ijk},}
with all other components being zero.
Taking  $f_{ijk}$ to be the structure functions for $SU(2) \times SU(2)$
and moving the $D0$ branes in the $X^9$ direction, one finds that the 
$D0$ branes have puffed up into a fuzzy $S^2 \times S^2$, with a radius
which is again linearly dependent on the velocity.
Similarly, choosing $f_{ijk}$ to be the structure functions 
for $SU(3)$, one finds, depending on the choice of irreducible representation,
fuzzy $CP^2$ or $SU(3)/(U(1) \times U(1))$.  
In all these cases there is  also a description in terms of a higher dimension
expanded brane; a $D4$-brane for $S^2 \times S^2$ and $CP^2$ and a 
six-brane for $SU(3)/(U(1) \times U(1))$. These expanded branes have 
a world volume $U(1)$ gauge field turned on and carry the same quantum numbers as the 
puffed up $D0$ branes \TV. They also have induced electric and magnetic multipole moments.
The $S^2 \times S^2$ configuration has electric and magnetic quadropole moments,
while $CP^2$ and $SU(3)/(U(1) \times U(1))$ have  electric and magnetic 
dipole moments.

\newsec{Giant Gravitons in M theory}

$D0$-branes in M-theory are gravitons moving along the $ M$ direction.
The example we considered above for the magnetic case can be
interpreted in M theory as a situation where the gravitons move both
along the $M$ direction and along $X^4$. In fact the action \acmagdtwo\
and hence the Hamiltonian \hamtwo\ are precisely those of a M2 brane
with a momentum $P_M = {N \over gl_s}$ in the M direction 
(where $g$ is the string coupling).

It is natural to assume that
the simpler situation where the graviton moves along say only the $M$
direction with $F^{(4)}_{123M}$ turned on would also result in the
graviton expanding into a transverse sphere. The $M2$-brane in turn
should be transverse to and moving along the $M$ direction and should
be expanded along the classical two-sphere.

We will  see next that this is indeed true.
To keep the description for the graviton and two-brane under control
we analyze this in the Type IIA limit first. 
In Type IIA one is looking for a solution consisting of  $N$ static $D0$ branes 
subject to an external $H_{123}$ field. 
In fact this situation was considered in \M.
The energy  for this static configuration  is
\eqn\ach{E=N T_0 
-{T_0 \over 4 \lambda^2} \sum_{ab}Tr([X^a,X^b]^2) -i {T_0 \over 3 \lambda}
 H_{ijk} Tr(X^iX^jX^k).}
Setting,
\eqn\defh{H_{123}=-F,}
(with all other components, not related by symmetries, equal to zero) 
in \ach\ one sees that this is 
 in fact identical to \act\ above. Thus the resulting solution which minimises the 
energy is a  non-commutative two sphere and the 
$D0$ branes (equivalently the M-theory
graviton) are  indeed puffed up in the presence of the external field. 
The radius and energy of the configuration is given by \charele.

To analyze this situation from the two-brane point of view we use the description in
terms of the $D2$ brane action. As was mentioned above the $M2$ brane is expected to be 
transverse to the $M$ direction. In this case, the $D2$ and $M2$ brane world volume theories 
are related by a duality transformation which turns the 
scalar field  corresponding to the $M$ direction in the $M2$-brane world 
volume theory, into the $D2$ brane gauge field. 
Thus we expect the $D2$ brane theory to have $N$ units of magnetic flux.
 The energy  for a static $D2$-brane with $H_{123}$ turned on is
\eqn\achdtwo{E=4 \pi T_2\sqrt{r^4 + ({N\lambda \over 2} -{1 \over 3}F r^3)^2}.} 
This is  different from \eleex. The radius of this brane configuration 
can be obtained by minimising \achdtwo\ with respect to $r$.
Note that in \achdtwo\ unlike \eleex, the energy grows as $r^3$ for large $r$ so 
 a minimum exists for all values of $F$.  In general the energy and radius
we obtain will not agree with \charele. However, in the limit when 
$r^2  \ll N \lambda $ the square root can be expanded in \achdtwo\
and once again yields the three terms of \expaact. Thus the energy and radius 
agree with \charele. 

There is one important issue to be noted in our discussion above. We have neglected the 
backreaction of $H_{123}$ on the metric. 
We leave a full discussion after including backreaction effects for the future 
and content ourselves here with some estimates. 
The metric perturbation induced  by \defh\ over a region of size $R$
is
\eqn\metrest{\delta h_{\mu \nu} \sim R^2 F^2.}
The same estimate also applies for the dilaton.  It is useful to start in the limit
when the back-reaction is small, i.e.,
\eqn\back{R^2 F^2 \ll 1.}
 In this limit one can make a self-consistent 
estimate and argue that 
\eqn\estrad{R \sim \lambda F N,}
 this is the same order of magnitude as 
 \elecr. The argument goes as follows. Let us assume that 
\estrad\ is true. Then from \ach\ and \estrad\ one can argue that the 
the leading order contribution to the energy  is  
\eqn\leade{E_0=NT_0.}
 The first corrections to this is  of order 
\eqn\firste{E_1 \sim T_0 \lambda^2 F^4 N^3.}
The leading contributions from the second and third terms in \ach\ 
are of this order and the back-reaction can be neglected
in obtaining them. However, the back-reaction is important in obtaining the 
subleading contribution from the first term in \ach.
 This term depends on the 
dilaton through the $D0$-brane tension. Taking into account the back-reaction
in the dilaton of order \metrest\  yields its contribution, which is also 
of order \firste.
The resulting three of order \firste\  must be minimised
to yield a radius. One expects an answer of order \estrad, since 
all three terms are then comparable. 

From the expanded $D2$-brane point of view, note that \estrad\ and 
\back\ imply that $R^2 \ll N \lambda$. 
So  the limit when the back reaction is small, is precisely the limit 
discussed above when the square root in \achdtwo\ can be expanded resulting 
in three terms which correspond to \ach. When the backreaction is neglected,
we showed above that these three terms agree quantitatively with
those in \ach. When the backreaction is included one can show that 
a similar agreement persists in the limit \estrad, \back.

\newsec{Giant Gravitons in Brane backgrounds}

We have seen above that M theory gravitons in appropriate background fields 
can turn into expanded branes. The phenonemon was investigated 
in AdS space in \GS, as was mentioned in the introduction.
It was found that the Hamiltonian of a $p$ brane moving
on the $p+2$ sphere  of a $AdS_m \times S^{p+2}$ space-time 
(and not performing any other kind of motion or oscillation)
is {\it exactly}
the same as that of a massless particle with the same quantum numbers.
It is rather remarkable that a ``heavy'' object like a brane can have
an energy with a gapless spectrum. The reason behind this is a delicate
cancellation of the effect of the brane tension with the energy due to
coupling to the background $F_{p+2}$ form gauge field leading to a BPS
like condition \djm. Furthermore the configurations
which saturate this BPS bound also preserve half of the supersymmetries
of the background \MG.
This mechanism seems to depend on the details of the
background geometry and makes one wonder whether it is a phenomenon
restricted to $AdS_m \times S^{p+2}$ spacetimes. 

In this section we will show that blowing up of gravitons into expanded branes
 with the same energy is  much more general and  
occurs in a wide variety of spacetimes.
The backgrounds we consider are those of both extremal and non-extremal branes.
 Significantly this includes backgrounds with no supersymmetry.

\subsec{$Dp$ branes in background of $N$ ~$D(6-p)$ branes}

To keep the discussion general we consider a $(6-p)$ brane geometry with
a metric of the form:
\eqn\metra{ds^2=-g_{tt}dt^2+\sum_{i=1}^{6-p}g_{ii} (dX^i)^2 +g_{rr}dr^2  +
f(r) r^2 d \Omega_{p+2}^2.}  
This metric has an $SO(p+2) $ rotational
symmetry and we will in particular be interested in states which carry
$SO(p+2)$ angular momentum.  In the discussion below it will be useful
to choose the following coordinates on a unit $p+2$ sphere
\foot{The considerations below are valid for all $p > 0$. The case of
$p=0$ is discussed separately at the end of the subsection}:
\eqn\coordunit{d\Omega_{p+2}^2={1\over 1-\rho^2}d\rho^2+
(1-\rho^2)d\phi^2+\rho^2 d\Omega_p^2,} 
where, $d\Omega_p^2$, refers to
the standard metric of the $S^p$ sphere, which we take to be
parametrised by the angles $\theta_1, \theta_2, .... \theta_{p-1},
\psi$ with $0\leq \theta_i\leq \pi$ and $0 \leq \psi \leq 2\pi$.

Following \GS\ we now consider configurations in which the $p$-brane
wraps the $S^p$ sphere. We choose a static gauge where the time parameter
of the worldvolume $\tau = t$ while the $p$ angular spacelike parameters
$\sigma_i$ are set to be equal to the angles on the $S^p$, $\sigma_i
= \theta_i$. The dynamical variables are then $r(t,\theta_i), X^i
(t, \theta_i), \rho (t,\theta_i)$ and $\phi (t,\theta_i)$. We consider
configurations where these quantities do not depend on $\theta_i$ so that
there are no brane oscillations. Furthermore since there is complete
translational symmetry along $X^i$ the corresponding momenta are conserved.
We will study motions where these momenta are identically zero. Motions
with nonzero momenta along the brane can be easily obtained by performing
boosts. In our ansatz then, the dynamical variables are $r(t), \rho(t)$ and
$\phi(t)$. The DBI action is
\eqn\pdbi{S_{DBI} = -T_p V_{p}\int dt e^{-\phi} (f(r) \rho^2 r^2)^{p/2}
\sqrt{ g_{tt}-g_{rr}{\dot {r}}^2-g_{\rho\rho}{\dot {\rho}}^2 - 
g_{\phi \phi}{\dot {\phi}}^2},}
where $V_{p}$ stands for the volume of the $p$-sphere and we have carried out the 
integrals along the $S^p$ world volume directions. 

In addition the brane action gets a contribution from the  Cherns Simon term. This arises because the 
$(6-p)$ brane gives rise to a magnetic $p+2$ form field strength 
(or equivalently an
electric $F^{(8-p)}_{013 ...(6-p)r}$ field strength)  that 
threads the $p+2$ sphere in \metra.  It is:
\eqn\formfield{T_p F_{\rho\phi\theta_1 ..\theta_{p-1}\psi}= 
{2 \pi N \over V_{p+2}} \rho^p 
\epsilon_{\theta_1 ... \theta_{p-1} \psi},}
where $\epsilon_{\theta_1 ... \theta_{p-1} \psi}$ is the volume form of the 
unit $p$-sphere and   $V_{p+2}$ denotes the
total volume   of the unit $p+2$ sphere respectively.
$N$ in \formfield\ refers to the number of $(6-p)$ branes.  
From \formfield\ we see that with an appropriate choice of gauge we can take
\eqn\gaugepot{T_p C_{\phi\theta_1 ...\theta_{p-1}\psi}=
{2 \pi N \over V_{p+2} (p+1)} \rho^{(p+1) } \epsilon_{\theta_1 ... \theta_{p-1} \psi},}
where $\epsilon_{\theta_1 ... \theta_{p-i}}$ is the volume form on a unit
$S^{p}$ sphere. 

The Cherns Simon term (after integrating over the $S^p$ world volume
directions again) is given by
\eqn\csterm{S_{CS}={2 \pi V_p N \over V_{p+2} (p+1)}  \int dt \rho^{p+1} 
{\dot \phi}. }
Now using  the fact that 
\eqn\valvp{V_{p}={ 2 \pi^{(p+1) \over 2} \over \Gamma({p+1 \over 2})},}
we get the full action 
 from \pdbi\ and \csterm\ to be  
\eqn\fullact{S=-T_p V_{p}\int dt e^{-\phi} (f(r) \rho^2 r^2)^{p/2}
\sqrt{g_{tt}-g_{rr}{\dot r}^2-g_{\rho\rho}{\dot \rho}^2 - g_{\phi \phi}{\dot \phi}^2 }+
N \int dt  \rho^{p+1} {\dot \phi}. }
To study the resulting dynamics it is useful to construct the Hamiltonian for this 
system. 

The momenta are:
\eqn\momp{\eqalign{P_r={\partial L  \over \partial {\dot r}} =& {T_p V_p e^{-\phi} 
\over \sqrt{g_{tt}-g_{rr}{\dot r}^2-g_{\rho\rho}{\dot \rho}^2 - g_{\phi \phi}{\dot \phi}^2 }  }
  (f \rho^2 r^2)^{p/2} g_{rr} {\dot r} \cr
P_{\rho}={\partial L  \over \partial {\dot \rho}} =&{T_p V_p e^{-\phi}
\over \sqrt{g_{tt}-g_{rr}{\dot r}^2-g_{\rho\rho}{\dot \rho}^2 - g_{\phi \phi}{\dot \phi}^2}}    
            (f \rho^2 r^2)^{p/2} g_{\rho\rho} {\dot \rho} \cr
P_{\phi}={\partial L  \over \partial {\dot \phi}}=&{T_p V_p e^{-\phi}
\over \sqrt{g_{tt}-g_{rr}{\dot r}^2-g_{\rho\rho}{\dot \rho}^2 - g_{\phi \phi}{\dot \phi}^2} } 
              (f \rho^2 r^2)^{p/2} g_{\phi \phi} {\dot \phi} + N \rho^{p+1}. }}

The Hamiltonian then is 
\eqn\hamdp{\eqalign{H=&P_r{\dot r} + P_{\rho}{\dot \rho} + P_{\phi}{\dot \phi} - L \cr
                     =&\sqrt{g_{tt}} \bigl [ (T_p e^{-\phi} V_p)^2 (f(r) \rho^2 r^2)^{p} 
+ {P_r^2 \over g_{rr}} + {P_{\rho}^2 \over g_{\rho\rho}}+
{(P_{\phi}-N\rho^{p+1})^2 \over g_{\phi\phi}} \bigr ]^{1/2}. }}

Now notice that if
\eqn\impcond{T_p e^{-\phi} V_p (f(r) r^2)^{{p+1 \over 2}} = N,}
the first and last terms within the square brackets above can be combined,
exactly as in \djm\ 
and the Hamiltonian can be rewritten as
\eqn\hamre{H=\sqrt{g_{tt}} \bigl [ {P_{\phi}^2 \over f(r) r^2} 
+ {P_r^2 \over g_{rr}}
+{P_{\rho}^2 \over g_{\rho\rho}}+ {(\rho P_{\phi}-N\rho^p)^2 \over 
g_{\phi\phi}} \bigr ]^{1/2}, }
where we have used from \metra\ and \coordunit\ that 
$g_{\phi\phi}=f(r) r^2 (1-\rho^2)$.
\impcond\ is an important condition and we will refer to it 
repeatedly in the discussion below. 

$P_{\phi}$ is a constant of motion. It is clear from \hamre\ that for
a given $P_\phi$ the lowest energy configuration satisfies
$P_{\rho}=0$ for all time. This is because $\rho$ does not appear in
the first two terms and the equation of motion for such configurations
simply require that the last term vanishes. This gives the equilibrium
value of $\rho$ in terms of $P_\phi$
\eqn\condsize{P_{\phi}=N\rho^{p-1},}
a condition independent of $r$.
The Hamiltonian then reduces to
\eqn\redham{H= \sqrt{g_{tt}} \bigl [ {P_{\phi}^2 \over f(r) r^2} + 
{P_r^2 \over g_{rr}} \bigr ]^{1/2}.}

Now, we come to the punch line of this section. Notice, that \redham\ is exactly the
Hamiltonian for a massless particle which carries angular momentum
$P_{\phi}$ on the $S^{p+2}$ sphere - one simple way to see this is to
consider the Laplacian in the WKB approximation.  Thus, as long as
\impcond\ is met, the expanded $p$-brane behaves like a massless
particle.

However, unlike a usual massless particle the brane has a bounded
angular momentum for such motions, just as in \GS. 
This follows from \condsize. Since
$ 0 < \rho < 1$ the maximum angular momentum is $N$. This is the
analog of the stringy exclusion principle.

It important to note that the physical size of the brane in the string
metric
\eqn\physize{ R = f^{1/2}r\rho}
depends on $r$ and hence is  not a constant of motion.
However, this is entirely due to the change in radius of the 
transverse $(p+2)$ sphere.

We now  examine specific examples to see when \impcond\ is met. 
First, consider the near horizon geometry for the  extremal $D(6-p)$-brane
\ref\HS{G. T. Horowitz and A. Strominger, Nucl. Phys. {\bf B 360},197-209 (1991).}.
In this case 
\eqn\metricp{\eqalign{g_{tt}=g_{ii}=& H^{-1/2}  \cr
                      g_{rr}=& H^{1/2} \cr f(r)=& H^{1/2} \cr
                      e^{\phi}=& H^{{p-3 \over 4}}, \cr {\rm where}
                      \quad H =& (R/r)^{p+1} \cr {\rm and } \quad
                      R^{p+1}= &2^{p-1}\pi^{{(p-1)\over 2}}\Gamma({p+1
                      \over 2}) g_s l_s^{p+1} N, }} 
where $N$ is the number of $D(6-p)$-branes.  We see then that \impcond\
is indeed met.  We would like to emphasize that for $p\ne 3$ the near
extremal geometry is not AdS.  Next consider the non-extremal, near
horizon geometry \ref\IM{N. Itzhaki, J. M. Maldacena, J. Sonnenschein and 
S. Yankielowicz, Phys.Rev. {\bf D58}:046004,1998, 
{\tt  hep-th/9802042}.}.  In this case $g_{tt},g_{rr}$ are different from
their values in the extremal case, but $f(r), e^{\phi}$, and $H$ are
still unchanged from \metricp\ so that once again \impcond\ is
met. This illustrates that even in non-supersymmetry preserving
backgrounds the expanded brane can behave like a massless particle.

Our conclusions above also apply to non-extremal, near horizon $M2$
and $M5$ brane metrics.  In fact, the discussion above can be carried
over to those cases almost directly.  Let us briefly sketch out how
this happens.  The total number of dimensions now is  eleven
and in \metra\ the sum over coordinates parallel to the brane goes
from $i=1$ to $(7-p)$. The metric has an $SO(p+2)$ symmetry.  The case
$p=5$ with $SO(7)$ symmetry refers to the $M2$-brane background while
case $p=2$ with $SO(4)$ symmetry to the $M5$-brane background.  The
dynamics of the $p$-brane moving in this background is described by an
action consisting of a BI term and a Cherns Simon term. The BI term is
given by \pdbi\ without the dilaton factor $e^{-\phi}$, where $T_p$
stands for the tension of the $p$-brane, while the CS term is still
\csterm, with $N$ being the number of $(7-p)$ branes.  The crucial
condition \impcond\ is now replaced by
\eqn\impcondm{T_p V_p (R_{AdS})^{p+1}=N.}
One can see that this is met for the $M2$ and $M5$ brane
geometries. Moreover, it is independent of whether we are dealing with
the extremal or non-extremal cases.  Thus  for both these cases the
special solutions which satisfy \condsize\ (with $p=5$ and $p=2$
respectively) yield an energy,
\eqn\blueham{H=\sqrt{g_{tt}} [ {P_{\phi}^2 \over R_{Ads}^2} 
+ ({r \over R_{AdS}})^{p+1 \over 2} P_r^2]^{1/2},} 
which is the same as that for a massless particle.

Two special cases $p=1$ and $p=0$ in the discussion above
are worth commenting on.  In the $p=1$ case,
\condsize\ is independent of $\rho$ and  the special solution,
with an energy equal to the massless case, exists only for 
\eqn\condone{P_\phi=N.}
Moreover, if \condone\ is true the potential for $\rho$ vanishes. 
Thus,  for this special value of angular momentum there is a one-parameter family
of solutions all of which behave like massless particle. 
These comments are equally valid in the extremal and non-extremal cases.
A closely related case is realised when one considers
a fundamental string in an $NS$ $5$-brane background.
In this case the near-horizon geometry has a three-sphere
of constant radius $R_3$. \impcond\ is replaced by  a similar condition 
which does not involve the dilaton and relates $R_3$ to the number of 
five-branes. The condition is in fact met leading again to 
special solutions for  \condone\ which behave like massless particles both
in the extremal and non-extremal cases.    

For $p=0$  the formulae above receive some  modifications.
First, the last term in the metric on $S^{p+2}$ in \coordunit\ is not
present. The coordinate $\rho$ can still be used \foot{Setting $\rho=\cos\theta$
yields the familiar metric on the two-sphere.}. 
Secondly in \csterm\ one has to define $V_0 = 1$. As a
result the coefficient of the Chern-Simons term in \fullact\ acquires
an additional factor of $1/2$. This modifies the condition \impcond\ to
\eqn\impcondzero{T_0 e^{-\phi} (f(r) r^2)^{{1 \over 2}} 
= {1\over 2}N~~~~~(p=0).}
One can verify \metricp\ that this is indeed met for the 
$6$-brane extremal and non-extremal, near-horizon metrics.  
This case, differs physically  from $p > 0$ in some important ways. Note that
for $p \geq 1$ the brane in question has {\it no net charge} since it
is always wrapped on a $S^p$ which is contained in the $S^{p+2}$.
However a single zero brane carries charge and consequently 
does not have the same quantum numbers as a graviton. One possiblility
is to consider a pair of a zero brane and an anti zero brane - this
would have a state with the same energy as a pair of gravitons. Furthermore
the relationship which determines the equilibrium value of $\rho$ is
pathological since it implies that there is a {\it lower} bound on 
the angular momentum. Some of these problems can be possibly resolved
by introducing a coupling between the $D0-{\bar D0}$ pair \GS.

Finally, consider for example the extremal $Dp$-brane geometry without
taking the near horizon limit. In this case the Harmonic function in
\metricp\ is replaced by $H=1+(R/r)^{p+1}$.  Now one can verify that
\impcond\ is no longer met. For example in the asymptotic region the
dilaton and $f(r)$ go to constant so the l.h.s of \impcond\ grows like
$r^{p+1}$. 
In fact one show in this case that the energy of the
expanded brane configuration is always bigger than than for a massless
particle. Furthermore it is no longer possible to obtain motions with
$P_\rho = 0$ with nonzero $\rho$, a fact which may be easily seen in
the deep asymptotic region of large $r$.

\subsec{$4D$ and $5D$ Black Holes}

The condition \impcond\ (or \impcondzero\ for $p=0$) is also satisfied
by appropriate branes moving in the near horizon geometries of five
and four dimensional black holes in string theory.  As in the previous
subsection we have to discuss the motion of branes which are {\it
magnetic duals} of the branes which produce the background.

Consider first the $5D$ extremal black hole obtained in IIB string
theory compactified on a $T^4 \times S^1$ with $Q_5~~D5$ branes
wrapping $T^4 \times S^1$ and $Q_1~~D1$ branes wrapping the $S^1$. The
magnetically dual branes are then, (i) a $D1$ brane wrapping a circle
on the transverse $S^3$ which can couple to the mangetic 3-form field
strength threading the $S^3$ and (ii) a $D5$ brane wrapped on $T^4$
and a circle on the $S^3$.  These are further wrapped on a circle on
the transverse $S^3$ and move on it.  Both cases relate to the $p=1$
case of the previous subsection with $N$ replaced by $Q_5$ for (i) and
$N$ replaced by $Q_1$ for case (ii).  Using the well known background
geometry \ref\bhref{See e.g. J. Maldacena, PhD. Thesis,
{\tt hep-th/9607235}} it is straightforward to check that
\impcond\ is satisfied. The only values of the angular momentum for
which one has equilibrium brane configurations with the same energy
as gravitons are $Q_5$ and $Q_1$ respectively. In this case the
geometry is in fact $AdS_3 \times S^3 \times T^4$ (which has been
considered in \GS) and the exclusion
principle bound is $Q_1Q_5$ which differs from both these values
\foot{ We would like to thank S.D. Mathur for discussions about this
point}.

Similarly the four dimensional black hole in IIA string theory
compactified on $T^4 \times S^1 \times {\tilde{S^1}}$ is made of $Q_2$~$D2$
branes wrapping $S^1 \times {\tilde{S^1}}$, $Q_6~~D6$ branes wrapping $T^4
\times S^1 \times {\tilde{S^1}}$ and $Q_5~~NS5$ branes wrapping $T^4 \times
{\tilde{S^1}}$. For extremal black holes the geometry is
$AdS_2 \times S^2 \times T^6$.
Now the magnetically dual objects are ; (i) $D4$ branes
wrapping the $T^4$ (ii) $D0$ branes, and (iii) $F1$ string wrapping
$S^1$. They all move on the transverse $S^2$.  All these relate to the
$p=0$ case discussed in the previous section with the coefficient of
the Chern Simons term replaced by $\half Q_2$ for (i), by $\half Q_6$
for (ii) and $\half Q_5$ for (iii). Once again it may be verified that
\impcondzero\ is satisfied for all the cases (for case (iii) the
dilaton factor is absent in \impcondzero, as commented above).

In both these cases the addition of a momentum along the $S^1$, or
addition of nonextremality does not change the result since they do not
affect the $S^3$ or $S^2$ parts of the metric respectively.

\subsec{Discussion}

It is worth discussing the results of the above calculation in some
detail.

Let us begin by relating the discussion of the previous section to \GS.
The discussion in \GS\ was for AdS space and overlaps with the
analysis above in the the $D3$-brane, $M2$ and $M5$-brane cases.  The
one difference is that we have used Poincare coordinates instead of
Global ones.  The Hamiltonian for the $M2$, $M5$ branes is given in
\blueham. For the $D3$ brane case we have from \redham\ and $p=3$
\eqn\pads{H={r \over R}[{P_\phi^2 \over R^2} + 
{R^2 \over r^2} P_r^2]^{1/2}.}
The prefactor $\sqrt{g_{tt}}$ in \blueham\ \pads\ is the usual
red-shift in energy.  Due to it we see that a massless particle, or
equivalently an expanded brane, initially at rest in the radial
direction will fall into the black hole.  In contrast, in global
coordinates a particle at the center of AdS does not move and the
energy in global coordinates is equal (in units of the radius) to the
angular momentum, making the BPS nature of the state more transparent.

For extremal D-brane backgrounds, the equation of motion which follows
from the Hamiltonian \redham\ can be written as
\eqn\pbranemotion{ ({\dot r})^2 + U(r,P_\phi,E) = 0}
where $E$ is the energy and 
\eqn\fakepotential{U(r,P_\phi,E) = {P_\phi^2 \over E^2}~r^{2p}
- r^{p+1}}
It is thus clear that motion is always restricted between $r=0$ and
a turning point 
\eqn\turn{ r_t = ({E \over P_\phi})^{{2 \over (p-1)}}}
So far as motion in the radial direction is concerned the angular
momentum provides a potential well which prevents the particle to
escape to large $r$. For nonextremal near-horizon D-brane backgrounds,
however, the angular momentum provides a finite potential {\it barrier}
near the horizon, just as in the vicinity of Schwarzschild black holes.

In the non-AdS extremal backgrounds discussed above, the expanded $p$ brane
solution is not supersymmetric, as best as we can tell.  
Certainly this is true
in the non-extremal geometries.  Despite this one finds that the
expanded branes behave like massless particles as long as \impcond\ or
\impcondm\ is met.  Unfortunately, we do not understand the
significance of this condition well enough at the moment. One comment
is worth making in this context though.  Consider the $Dp$-brane case
first. Due to the factor $e^{-\phi}$ multiplying the BI action, the
metric seen by the $Dp$-brane differs from the string metric by a
conformal factor. It is
\eqn\conref{ds^2_{p}=(e^{-2 \phi})^{1 \over p+1} \ \  ds^2_{string}.}
Interestingly, the resulting metric is AdS space.  
The $Dp$-brane in the course of its motion sweeps out a $p+1$
dimensional surface of the topology $S^1 \times S^p$. \impcond\ sets
the volume of this surface in units of the $p$ brane tension to  
equal to $2 \pi N$, where
$N$ is the total $p+2$ form flux threading the $S^{p+2}$.
Alternatively, perhaps the more useful way to state \impcond\ is that
the radius of the $p+2$ sphere, in the $p$-brane metric, must be a
constant and determined by the magnetic flux.  For the $M2$, $M5$
brane cases no rescaling is required and the $p$-brane metric is the
$M$ theory metric. \impcondm\ then sets the volume of the $S^1 \times
S^p$ surface equal to $2 \pi N$ in  units of the $p$ brane tension, or alternatively the
radius of the $p+2$ sphere equal to an appropriate constant.

We noted before that the equilibrium size of the $p$-brane {\it in the
string metric} is not a constant of motion. The above considerations,
however, show that the equilibrium size {\it in the $p$-brane metric}
is indeed a constant of motion.

Physically, it seems puzzling that an extended brane configuration
manages to have the same energy as a massless particle.
The answer lies in the fact that the expanded $p$-brane is
the magnetic dual of the $(6-p)$ which gives rise to the background.
The resulting Cherns Simon coupling  reduces the energy required, for fixed 
angular momentum, by just the right amount to equal the extra 
potential energy needed to support the extended brane.

The special solutions \condsize, \redham, \blueham, exist only when
the angular momentum is less than $N$.  For higher angular momenta, it
is safe to conjecture that there is no expanded brane configuration
which behaves like a massless particle. This is the analogue of the 
stringy exclusion principle.

We now turn to examining two more issues in some detail.  In the next
section we show that the graviton and the expanded brane descriptions
are valid for different and non-overlapping ranges of angular
momentum. This is important if expanded branes are to be identified
with gravitons.  In the last section of the paper we focus on one
special instance of the general discussion above: the $p=2$ case. In
this case we have a $D2$ brane expanded into a two-sphere in the $D4$
brane background.  One can in addition turn on $N_0$ units of magnetic
flux on the world volume of the $D2$ brane. We show that for an
appropriate region of parameter space this configuration can be
described as $N_0$ $D0$ branes, puffed up into a non-commutative
two-sphere, and rotating on the $S^4$.

\newsec{Gravitons vs Expanded Branes}

In general one would expect that the descriptions in terms of a graviton and 
an expanded brane state are vaild in different regions of parameter space. 
Certainly one can argue this for the
AdS backgrounds studied in \GS\ where the BPS nature of the states ensures 
there cannot
be multiple copies. But even more generally for the $p$-brane extremal and 
non-extremal backgrounds one expects only one of the two  description to be 
valid.

We will now argue that this is indeed the case. The graviton and
expanded brane description are valid for different  
values of angular momentum \foot{Here angular momentum refers to
rotations on $SO(p+2)$.}.

Let us start with the case of an $AdS_m \times S^{p+2}$
background in $M$ theory. 
In analyzing the graviton states
one can think of doing a Kaluza Klein reduction on the $S^{p+2}$.
The graviton then turns into a massive state with mass
\eqn\massga{M\sim P_{\phi}/R_{AdS} \sim  M_{Pl} {P_{\phi} \over  N^{1/(p+1)}},}
where $P_{\phi}$ refers to the angular momentum, and we have used the
fact that
\eqn\defr{R_{AdS} \sim {N^{1/(p+1)}\over M_{Pl}}.}
In order to neglect the higher derivative terms in the action, 
arising for example 
from higher powers of the curvature, 
and treat the graviton in a controlled manner we need
\eqn\cconda{M \ll M_{Pl},}
leading to 
\eqn\ccondba{P_{\phi} \ll N^{1/(p+1)}.}
The alternative description in this case involves an expanded $p$-brane.
This description in under control when 
the brane has a big size  compared to the Planck Scale
 so that acceleration terms can be neglected
and one can work with the BI $+$ CS action.
This gives a condition
\eqn\ccondb{R_{AdS}\  \rho \gg 1/M_{Pl}.}
Substituting for $\rho$ from \condsize\   we get 
\eqn\ccondc{P_{\phi}\ \gg N^{2 / p+1}.}
We see that \cconda\ and \ccondc\ can never be simultaneously met.

Before proceeding let us make two comments. First, our use of the word
 graviton should not be taken literally. We simply mean a fluctuation 
about the $AdS_m \times S^{p+2}$ supergravity 
background which is massless in  $11$ dim. 
Second, the Planck scale in \cconda\ is the $11$ dim. Planck scale. 
The gravitational backreaction  after Kaluza Klein reduction
is governed by the $(9-p)$ dimensional Planck scale, which is bigger than $M_{Pl}$
since $R_{AdS} M_{Pl} \gg 1$. Requiring these to be under control, therefore,
is a less stringent condition than \cconda.

The $D3$ brane case is similar to the case above with the string scale 
playing the role 
of $M_{Pl}$. The general
$Dp$-brane case   has one new aspect: 
the radius of the $S^{p+2}$
is not constant in these cases. Inspite of this the argument above essentially
goes through.  
Consider a massless particle  moving on the $S^{p+2}$.
Carrying out a Kaluza Klein reduction on the $p+2$ sphere and demanding that
the resulting mass is smaller than the string scale yields the condition
\eqn\dconda{P_{\phi} \ll R_{p+2}/l_s,}
where $R_{p+2}$ is the radius of the $p+2$ sphere.
On the other hand for the expanded $Dp$-brane description to be valid
we have
\eqn\dcondb{R_{p+2} \rho = R_{p+2} ({P_{\phi} \over N})^{ 1\over p-1} \gg l_s.}
One can show that \dconda\ and \dcondb\ cannot be simulateneously valid if 
the dilaton $e^{\phi} \ll 1$ and string loop corrections to the 
supergravity description are under control. 

To see this note that 
\eqn\setcon{R_{p+2}=({R \over r})^{p+1 \over 4} r = 
({R \over r})^{p-3 \over 4} R .} 
So that the dilaton, \metricp,  can be expressed as
\eqn\setcondil{e^{\phi}=({R \over r})^{(p+1)(p-3) \over 4}=({R_{p+2} \over R})^{p+1}.}
Using the fact that $ R^{p+1} \sim g_s N l_s^{p+1}$ one can also express this as 
\eqn\conddila{e^{\phi}\sim {1 \over g_s N} ({R_{p+2} \over l_s})^{p+1 }.}

Now, if \dconda\ and \dcondb\ are simulateneously valid,
\eqn\dcondc{({R_{p+2} \over l_s}) \gg N ({l_s \over R_{p+2}})^{p-1}.}
But then it follows from \conddila\ that  
\eqn\dcone{e^{\phi}  \gg ({1 \over g_s N})
N^{p+1 \over p} = (1/g_s) N^{1/p} \gg 1,}
where the last inequality arises because $g_s \rightarrow 0$ and $N \rightarrow \infty$.
Thus in conclusion, when the supergravity approximation is valid,
the graviton and expanded brane description are never simultaneously valid. 

Let us comment on condition \dcondb\ in some more detail. 
Since $R_{p+2}$ depends on $r$ the massless particle 
after KK reduction gets a position dependent mass. In other words, in the 
KK reduced theory the particle satisfies a wave equations with a potential energy term. 
If this potential energy is of order the $M_{Pl}$ higher derivative terms 
will be important leading to \dcondb.

In summary then,  we have seen above that the
 massless particle description  and the expanded brane description are valid 
for different values  of the angular momentum. 
 As the rotational energy for the graviton increases
and becomes larger than the string scale (or Planck scale in $M$ theory) the
gravitons turn into an expanded brane configuration.
This is made all the more plausible by the fact that in several cases even without
supersymmetry the expanded brane solutions has the same energy, for fixed angular
 momentum, as the massless particle.  Once we accept this identification it can be 
extended to other cases,
where the expanded brane has a different energy from the massless particle.
For example, one can consider the expanded brane moving in the   full $(6-p)$ 
brane geometry.
Close to the horizon it behaves like a massless particle, but the identification should
still be valid as it moves further away.

\newsec{ Puffed Rotating  $D0$ branes}
In this section we return to considering one special case of the
general discussion in section 4:
a $D2$ brane moving in the background of the $D4$ brane.  This corresponds to
$p=2$;  the
background geometry has a $SO(4)$ rotational symmetry in this case.
We showed in section 4 that when the $D2$ brane carries $SO(4)$
angular momentum there is a particular solution \condsize,  for which
it behaves, in effect, like a massless particle \redham, and should be identified
with a supergravity mode. The $D2$-brane in  this configuration expands into
a two-sphere. Here we consider what happens when  in addition
$N_0$ units of magnetic flux are turned on in the world volume of the
$D2$ brane.  Through the usual Cherns Simons coupling it then acquires
$N_0$ units of $D0$-brane charge. We will see below that there is
another solution consisting of $N_0$,  $D0$ branes, also carrying the same
angular momentum, in which the  $D0$-branes
have  puffed up into a non-commutative two-sphere.
Thus, we have another example
of the Magnetic Moment effect discussed in section 2.2, but this time
in a non-constant four-form field  generated by  a $D4$-brane background.

\subsec{$D2$-brane with $U(1)$ flux}

To keep the discussion simple, we focus on the near-horizon extremal
$D4$-brane background. This is given by the metric and dilaton:
\eqn\metrdf{\eqalign{ds^2=&H^{-1/2}\bigl (-dt^2 + \sum_{i=1}^4 (dX^i)^2 \bigr )
 + H^{1/2} \sum_{i=5}^9 (dX^i)^2 \cr
& e^{\phi}=H^{-1/4}. }}
Here $X^i, i=5, \cdots 9$ denote the five transverse coordinates,
$r^2=\sum_{i=5}^9(X^i)^2$ and   $H=(R/r)^3$.
 To relate this to the metric \metra\ \coordunit\
we need the following relations:
\eqn\relcoord{\eqalign{X^5=r\sqrt{1-\rho^2} \cos\phi \  & \  X^6=r\sqrt{1-\rho^2} \sin\phi \cr
X^7=r \rho \cos\theta \ & \  X^8=r\rho   \sin\theta \sin\psi \cr
  X^9=r\rho \sin\theta \cos\psi. \   }}
This gives rise to the metric
\eqn\metricd{\eqalign{ds^2=H^{-1/2} & (-dt^2+(dX^i)^2) \cr
&+H^{1/2}\bigl [dr^2 + {r^2 \over 1-\rho^2}
d\rho^2+ r^2(1-\rho^2)d\phi^2+r^2\rho^2d\theta^2 + r^2 \rho^2\sin^2\theta d\psi^2
\bigr ], }}
which agrees with \metra, \coordunit.

With $N_0$ units of magnetic flux the $DBI$ action for the $D2$-brane, \pdbi,
is replaced by
\eqn\actmag{S_{DBI}=-T_2 4 \pi \int dt e^{-\phi} [ (H^{1/2}\rho^2 r^2)^2 +
{N_0^2 \lambda^2 \over 4}]^{1/2}
 \sqrt{ g_{tt}-g_{rr}({\dot {r}})^2-g_{\rho\rho}({\dot {\rho}})^2
 - g_{\phi \phi} {\dot {\phi}}^2 }}

The Cherns Simon term involving the coupling to the four form
is left unaltered, \csterm. For studying the dynamics,
most of the  discussion, \pdbi\ - \momp, can be carried over with only
slight modifications.
Once again focusing on the special case when $P_\rho=0$ and
\condsize\ is met yields a Hamiltonian :
\eqn\magham{H=\sqrt{g_{tt}} \bigl [\{4 \pi T_2 e^{-\phi} ( {N_0 \lambda \over 2}) \}^2
+{P_{\phi}^2 \over H^{1/2} r^2} + {P_r^2 \over g_{rr}} \bigr ]^{1/2}.}

The reader will notice that this is the Hamiltonian for a  particle
moving in the four-brane background with a with a position dependent mass
\eqn\pmass{m=2 \pi T_2 \lambda N_0 e^{-\phi}.}
Using the relation $2\pi T_2 \lambda = T_0$,  we see that
this mass is identical to that of $N_0$ $D_0$ branes. So the
expanded $D2$-brane solution with $N_0$ units of magnetic flux has an
energy exactly equal to $N_0$ $D0$-branes executing only center of mass motion, with
no relative displacement.
However, our experience in other situations discussed above would make
us suspect that there is another solution for $N_0$ $D0$-branes involving
a non-commutative two-sphere and this is the solution to be identified with the
expanded $D2$-brane \foot{For example the expanded $D2$ brane configuration has 
dipole moment with respect to $F^{4}$ as does the puffed up $D0$ brane configuration but
not the $D0$ brane configuration with no relative displacement.}.

\subsec{Puffed Rotating $D0$-branes.}

This expectation is indeed correct. To verify it  we need to consider the
Non-Abelian $D0$ brane Lagrangian  in the curved $D4$-brane geometry.
The Abelian BI Lagrangian, in static gauge, is given by:
\eqn\dab{L=-T_0N_0e^{-\phi}\sqrt{g_{tt}-g_{ij}{\dot x}^i {\dot x}^j }.}
This suggests that the  Non-Abelian Lagrangian (upto quartic terms)
  is given \foot{We note that this Lagrangian was also considered in \taylortwo.} by
\eqn\dzc{\eqalign{L=&-T_0  Tr \bigl [ e^{-\phi(X)}\sqrt{g_{tt}}
\{1 - {1 \over 2} { g_{ij}(X) \over g_{tt}(X)}
{\dot X}^i {\dot X}^j- {1 \over 4 \lambda^2}  \sum_{ab}
[X^a,X^b][X^c,X^d] g_{ac}(X) g_{bd}(X) \}  \bigr ] \cr
& + i{T_0 \over  \lambda} Tr[C^{3}_{ijk}(X) X^i X^j
{\dot X}^k ].}}
Notice since the background is space-time dependent the background fields
lie within the matrix traces above. The last term in \dzc\ is a Cherns Simon coupling
 which   arises as discussed in
\M\  and involves the RR $3$ form  gauge potential $C^3 $.
One big  difference between the discussion here and in   section 2.2 is
that the four form field strength is not constant. Consequently we need to work with
the full gauge potential $C^3$ in \dzc\ rather than its expansion to linear order.

Motivated by the $D2$-brane solution
discussed above and in section 4 we consider the following ansatz for
$D0$-brane solution
\eqn\ansatz{\eqalign{X^5=r\sqrt{1-\rho^2}\cos\phi \  \identity \  &
\  X^6=r\sqrt{1-\rho^2}
\sin\phi \  \identity \cr
X^{i+6}=& {2 \over N_0} r\rho  \ J^i, \{i=1,2,3\},  }}
where $J^i$ stand for $SU(2)$ generators in the  $N_0$ dimensional irreducible
representation.
Further, we take $r, \phi$ to be time dependent and take $\rho$ to be time
independent.
All the other coordinates, parallel to the $4$-brane
 are taken to be  a constant multiple of the identity.
It is worth pointing out that   the coordinates $X^{7},X^{8},X{^9}$ do not
commute and form
a non-commutative two sphere;
further,
\eqn\comm{\eqalign{(X^7)^2+(X^8)^2+(X^9)^2=&r^2\rho^2\identity, \cr
(X^5)^2+ (X^6)^2 +(X^7)^2+(X^8)^2+(X^9)^2=&r^2\identity. }} 

Now notice that the metric coefficients and the dilaton  dependence 
 in  \metrdf\
are a function of $r$ alone. In the Lagrangian \dzc\  $r^2$ is to be  replaced by
\eqn\repr{r^2 \rightarrow \sum_{i=5}^9(X^i)^2.}
Luckily,  due to \comm\  this is a multiple of the identity
matrix and can be taken out of the trace  and replaced  by the $c$-number $r^2$.
Thus we can take all the
dependence on the background  metric and dilaton
outside the matrix traces in \dzc. This leads to  considerable
simplification in evaluating the Lagrangian.

To evaluate the CS term  we need the three form potential
$C^3$ in the coordinates \metrdf.
$C^3$ in the coordinates \metra\ \coordunit\ was determined in  \gaugepot.
Using \relcoord\ to  change coordinates we 
get,
\eqn\cthree{\eqalign{T_0 C_{578}=-{ N \over 2} {1 \over r^3}{1\over \sqrt{r^2-r^2\rho^2}}
\sin \phi X^9 \ & \ T_0 C_{589}=-{ N \over 2}{1 \over r^3} {1\over \sqrt{r^2-r^2\rho^2}}
\sin\phi X^7 \cr
T_0 C_{597}=-{N \over 2}{1\over r^3}{1\over \sqrt{r^2-r^2\rho^2}}
\sin\phi X^8 \ &\
T_0 C_{678}={ N \over 2} {1 \over r^3}{1\over \sqrt{r^2-r^2\rho^2}} \cos\phi X^9 \cr
T_0 C_{689}={ N \over 2} {1 \over r^3} {1\over \sqrt{r^2-r^2\rho^2}} \cos\phi X^7 \  & \
T_0 C_{697}={ N \over 2} {1 \over r^3}{1\over \sqrt{r^2-r^2\rho^2}} \cos\phi X^8, }}
all other components are zero. We also remind the reader that $N$ in
\cthree\ refers to the number of $D4$ branes whereas $N_0$ stands for
 the number of $D_0$ branes. Once again, in
 the Chern Simon term, strictly speaking all  the space dependence in $C^3$ should be
 replaced by
functions of the coordinate matrices $X^i$. However due to \comm\
and the argument given above the $r$ and $r\rho$ dependence
can continue to be regarded as $c$ numbers. Similarly, since
$\phi$ can be expressed in terms of $X^5,X^6$ alone and  both of these  are
multiples of the identity it too can be regarded as a $c$ number.
This greatly simplifies the evaluation of the CS term. Each
component of $C^3$ now gives a term proportional to $Tr(X^7[X^8,X^9])$
which is proportional to the  simplectic two-form on the two-sphere.

Putting all this together finally yields a Lagrangian:
\eqn\finallag{L=-N_0T_0e^{-\phi}\sqrt{g_{tt}}[1- {1 \over 2}{g_{rr} \over g_{tt}}{\dot r}^2
-{1\over 2} {g_{\phi \phi} \over g_{tt}}
{\dot \phi}^2 +{2  \over N_0^2 \lambda^2} H r^4 \rho^4]
+ N {\dot \phi} \rho^3.}
To compare this with the $D2$ brane action we expand \actmag\  in the
non-relativistic limit and assume that the $N_0$ units of magnetic
field dominates the action compared to the surface   tension term.
One gets on keeping the leading term and the first correction
 (and after the identification $2\pi \lambda T_2=T_0$) exactly \finallag.
Since the the two Lagrangians agree, one can use our discussion in section 4 
for the $D2$-brane case to conclude again that minimizing with respect to 
$\rho$ yields the condition
\condsize\. Substituting this in the resulting  Hamiltonian yields: 
\eqn\nrham{H=\sqrt{g_{tt}} [N_0T_0e^{-\phi} +{1 \over 2}{1 \over N_0 T_0e^{-\phi}}
({P_\phi^2 \over H^{1/2} r^2} +{P_r^2 \over g_{rr}}) ].}

This  is the non-relativistic version of \magham\ and corresponds to
a non-relativistic particle of mass $N T_0 e^{-\phi}$ moving in the $D4$ brane
background.

In summary, we have found a solution to the Non-Abelian $D0$ brane action
in which the $D0$-branes rotating in the presence of the $D4$-brane background,
puffs up into a non-commutative two-sphere. The solution carries exactly
the same energy as if only the center of mass of the $D0$-branes was moving with
no relative displacement. There is also a expanded $D2$ brane solution
with the same quantum numbers and the same energy. 

\newsec{Acknowledgements} We would like to thank A. Dabholkar, A. Jevicki and
S. Mathur for discussions.

\listrefs
\end